 \documentstyle[aps,eqsecnum,preprint,epsfig]{revtex} 
\tighten 
\draft 
 
\newcommand{\cuo}{CuO$_2$\ } 
 
\begin{document} 
\preprint{\small To be published in {\it Physica A},
Ref: {\bf mu161/99043}
} 
% 
%--------------------------------------------------- 
% Creates wide abstract in twocolumn style 
%--------------------------------------------------- 
\ifpreprintsty 
\addtolength{\textwidth}{-1.5cm} 
\addtolength{\oddsidemargin}{-1.0cm} 
\addtolength{\evensidemargin}{-1.0cm} 
\else 
\twocolumn[\hsize\textwidth% 
\columnwidth\hsize\csname@twocolumnfalse\endcsname 
\fi 
%---------------------------------------------------- 
% 
\title{Simple model for the linear temperature dependence of the electrical resistivity of layered cuprates} 
\author{% 
Todor M. Mishonov% 
\thanks{Corresponding author: tel.:(++32)-16-327193,  fax.:(++32)-16-327983, 
e-mail:~todor.mishonov@fys.kuleuven.ac.be}% 
\thanks{On leave from: Department of Theoretical Physics, Faculty of Physics, University of Sofia, 5 J. Bourchier Blvd., Bg-1164 Sofia, Bulgaria,  
e-mail:~mishonov@rose.phys.uni-sofia.bg} 
and Mihail T. Mishonov% 
} 
\date{Received: June 3, 1999} 
\address{Laboratorium voor Vaste-Stoffysica en Magnetisme, Katholieke Universiteit Leuven,\\ Celestijnenlaan 200 D, B-3001 Leuven, Belgium} 
\maketitle 
\begin{abstract} 
The thermal fluctuations of  the electric field between the \cuo planes 
of layered cuprates are considered as an origin of the electrical resistivity. 
The model evaluation employs a set of separate plane capacitors each having area equal to the squared in-plane lattice constant $a_0^2.$ It is shown that the scattering of charge carriers by the fluctuation of electric charge in the conducting \cuo planes gives rise to the in-plane electrical resistivity $\rho_{ab}.$ Such a mechanism can be viewed as an analog of the Rayleigh's blue-sky law---the charge carriers are scattered by thermal fluctuations of electron density. 
\end{abstract} 
 
\ifpreprintsty\else\vskip1pc]\fi 
\narrowtext 
 
%-------------------------------------------------------------------------- 
\section{Introduction} 
\label{sec:intro} 
%-------------------------------------------------------------------------- 
The linear temperature dependence of the electrical resistivity 
$\rho_{ab} \propto T$ \cite{Gurvich,Martin,Ong} is one of the most important properties of the normal phase kinetics of high-$T_c$ layered cuprates\cite{Allen,Rebolledo}. 
However, despite the intensive investigations over a period of more than ten years and the numerous theoretical models proposed\cite{Maki,Ruvalds,Varma,Moshchalkov} this simple law does not yet have a unique explanation. In the physics of conventional metals it is well established that Pt, for instance, has a linear resistivity but the underlying physics is completely different. A parallel between similar behaviour for Cu and the data for La$_{1.825}$Sr$_{0.175}$CuO$_4$ and YBa$_2$Cu$_3$O$_7$ is available as well~\cite{Gurvich}. The linear temperature dependence of the resistivity is one of the most discussed problems in the physics of high-$T_c$ superconductors~\cite{Varlamov}. 
 
The aim of this paper is to present a simple model estimation explaining the  
$\rho_{ab} \propto T$ behaviour. Our model is based on the strong anisotropy of the electrical resistivity in layered cuprates. In the $c$-direction, perpendicular to the conducting \cuo planes ($ab$-planes), the electrical resistivity $\rho_c$ is significantly  lower than the in-plane one $\rho_{ab}$.  
For the $c$-polarised electric fields also the plasma frequency $\omega_c$ can be lower than the critical temperature 
\begin{equation} 
\hbar \omega_c < k_B T_c. 
\label{plasma} 
\end{equation} 
Plasma oscillations exist only in the superconducting phase. Plasmons in superconductors are observable only in the superconducting phase while in the normal phase they are overdamped. Therefore the criterion for applicability of the model is the lack of coherent transport in $c$-direction and almost frequency independent electromagnetic response for $\hbar \omega < k_B T$
(in Ref.~\cite{Abrikosov}, for instance, the author finds that coherence is not very important for the normal resistivity; for the applicability of our model it would suffice that only the mean free path in $c$-direction be not significantly larger than the lattice constant $c_0$ in the same direction).
In this case the electric field between the \cuo planes can be considered as a 
ilassical field and Boltzmann's distribution for the energy of the plane capacitors formed by these planes is applicable. Such a picture is probably most appropriate for YBa$_2$Cu$_3$O$_{7-\delta}$ and Bi$_2$Ca$_2$SrCu$_2$O$_8$ which contain double \cuo layers spaced by distance $d_0$ almost corresponding to the diameter of the oxygen ions. Every plaquete within a double \cuo plane is considered here as an independent plane capacitor with capacitance $C = \varepsilon_0 a_0^2/d_0,$ where $1/4\pi\varepsilon_0 \approx 9 \times 10^9$~Jm/C$^2,$ $\varepsilon_0$ being the dielectric permeability of vacuum and $a_0$---the lattice parameter of the \cuo plane; the distance between the copper and oxygen ions is thus given by ${1\over2}a_0.$  
% 
% 
%------------------------------------------------------------------------------ 
\section{Model} 
\label{sec:model} 
%------------------------------------------------------------------------------ 
 
The capacitors defined in Sec.~\ref{sec:intro} are the main ingredient of the proposed mechanism for creation of resistivity. According to the equipartition theorem their average energy can be written as ${1\over 2C}\langle Q^2\rangle  = {1\over 2} k_BT,$ which gives for the averaged square of the electric charge  
\begin{equation} 
\langle Q^2\rangle = \varepsilon_0 \frac{a_0^2}{d_0} k_B T. 
\label{Q2} 
\end{equation} 
 
Consider now the scattering of one nearly free charge carrier (e.g. a hole moving in a \cuo plane) by a localized charge $Q$. For definiteness the hole is assumed to move in the $x$-direction and pass by the charge $Q$ at minimum distance $r$ to it. The trajectory of the hole is approximated by a straight line and its velocity is nearly constant and corresponds in our model estimation to the Fermi velocity $v_F$.  
The time needed for the charge carrier to pass by the scatterer (the flypast-time) we evaluate as  
$\tau_Q \simeq 2r/v_F$~\cite{Orear}. The maximal Coulomb force acting perpendicular to the trajectory is $F_\perp=eQ/4\pi\varepsilon_0 r^2.$ 
Hence for the perpendicular momentum gained by the scattered hole one has  
$\Delta p_\perp \simeq \tau_Q F_\perp.$ 
The latter quantity is much smaller than the Fermi momentum $p_F=mv_F,$ 
and for small scattering angles $\theta_r\ll 1$ one has  
\begin{equation} 
\theta_r\simeq{\Delta p_\perp\over p_F} = {A\over r}, 
\end{equation} 
where $A\equiv eQ/4\pi\varepsilon_0E_F,$ $E_F = {1\over 2} m v_F^2$ is the Fermi energy, 
$m=m_{\rm eff}m_0$ is the effective mass in the \cuo plane, $m_{\rm eff}$ is the
dimensionless one, and $m_0$ is the free electron mass. 
We note that the Rutherford scattering is the same both in classical and quantum mechanics and the Coulomb logarithms in the theory of plasma exceed the accuracy of such model estimations. In principle the classical and quantum results might slightly differ upon taking into account prefactors containing Coulomb logarithms, however, the difference would be much smaller than the uncertainty introduced by the lack of knowledge of material and/or model parameters.   
Let us stress that any mechanism for electrical resistivity must incorporate in an essential way some mechanism for transmission of the electron quasimomentum to the lattice. The capacitor model does this implicitly. The thermally excited charges in the capacitors play the role of the defects in the metal.
Strictly speaking, the model considered is not purely electronic. It contains implicitly some weak inelastic electron-phonon interaction. In spite of its large relaxation time the latter ensures the equipartition theorem for the thermal energy of the independent capacitors.
 
The picture outlined above can be easily generalized to account for the influence of all scattering plaquetes along the $x=0$ line. Accordingly, the charge carrier travels at distance $r=\pm a_0, \pm 2a_0, \pm 3a_0,\dots$. Since the charges of the capacitors are independent random variables the average square of the scattering angle is an additive quantity~\cite{Migdal} 
\begin{equation} 
\langle \theta^2\rangle_{\rm line} = \sum_r \langle \theta_r^2\rangle=2\left({A\over a_0}\right)^2 
\left(1+{1 \over 2^2} + {1 \over 3^2} + \dots \right). 
\end{equation} 
The field outside the plane capacitors is essentially a dipole field, however, we will not discuss such details because the corresponding correction gives a factor of order of one: $\zeta(2) = 1 + 1/2^2 + 1/3^2 + \dots = \pi^2/6 \simeq 1.$ 
 
Now let us apply the discrete lattice model in order to address the diffusion of the charge carrier momentum on the Fermi surface. The mean free path $l$ is the distance after which the charge carrier "forgets" the direction of its earlier motion and scatters by $90^{\circ} = \pi/2$ having travelled a distance equal to $l/a_0$ lattice constants, {\it i.e.}
\begin{equation} 
\langle \theta^2\rangle_{l/a_0}= 
\left( {l\over a_0} \right) \langle \theta^2\rangle_{\rm line}  
= \left({\pi\over 2} \right)^2. 
\end{equation} 
Consequently, for the mean free path~\cite{note} we get finally  
\begin{equation} 
\label{path} 
l = 3\pi {4\pi \varepsilon_0 \over e^2}{E_F^2 d_0 \over k_BT}a_0. 
\end{equation} 

Inserting the transport life-time of the carriers defined for metals as $\tau_{\rm tr} = l/v_F$ in  Drude's formula for the conductivity 
\begin{equation} 
\sigma = {ne^2\tau_{\rm tr} \over m} = {1 \over \rho}, 
\end{equation} 
where $n$ is the number of charge carriers per unit volume and $e$ is the electron charge, one recovers the linear temperature dependence of the resistivity 
(distinguishing feature for the classical statistics, cf.~\cite{Emery}) 
\begin{equation} 
\rho(T) = {p_F \over ne^2l}  
= {p_F \over 3 \pi (4\pi\varepsilon_0) nd_0a_0 E_F^2} k_BT
= {k_BT \over 3 \pi^2\varepsilon_0 nd_0a_0 m v_F^3}. 
\label{rho}
\label{rho0}
\end{equation} 
It is remarkable that the squared electron charge $e^2$ is cancelled 
and $\hbar$ does not appear explicitly as well.
The electrical resistivity is by definition a property of the normal state whereas the cuprates
have attracted attention because of their high $T_c.$ That is why we consider it useful  to
perform a comparison with the experiment employing parameters of the superconducting phase. For clean superconductors when the mean free path $l(T_c)$ is much larger than the Ginzburg-Landau coherent length $\xi_{ab}(0)$ we can evaluate the effective 
mass $m_{\rm eff}$ as half of the effective mass of the Cooper pairs. 
For thin cuprates films, $d_{\rm film}\ll \lambda_{ab}(0),$ the effective mass of the Cooper pairs is determined by the electrostatic modulation of the kinetic inductance~\cite{Comment}, but in principle it is also accessible from the Bernoulli effect~\cite{Bernoulli},
the Doppler effect for plasmons~\cite{Doppler},
magnetoplasma resonances~\cite{magnetoplasma},
or the surface Hall effect~\cite{Hall}.
Further, the electron density can be extracted from the extrapolated to zero temperature in-plane penetration depth 
\begin{equation}
\frac{1}{\lambda_{ab}^2(0)}=  \frac{\mu_0 ne^2}{m}, \qquad
\mu_0=4\pi \times 10^{-7},  \qquad
\epsilon_0=1/\mu_0 c^2, \qquad
c =299792458 \; {\rm m s^{-1}}.
\end{equation}
The Fermi momentum $p_F$ can be determined on the basis of the model of a two-dimensional~(2D) electron gas which for bilayered cuprates gives 
\begin{equation}
n= \frac{2}{c_0} n^{\rm (2D)}, \qquad n^{\rm (2D)}=  \frac{p_F^2}{2\pi\hbar^2}.
\end{equation}
So after some elementary algebra the Eq.~(\ref{rho}) takes the form 
\begin{equation}
\frac{\sqrt{m_{\rm eff}}}{\lambda_{ab}^5(0)}\;
\frac{{\rm d} \rho}{{\rm d} T}
\approx C_{\rho \lambda}
\equiv {8\over 3 \pi^{1/2}} \frac{k_Be^5}{m_0^{1/2}(2\pi\hbar)^3}
\frac{\mu_0^{5/2}}{\epsilon_0}{1\over a_0 d_0 c_0^{3/2}} = {\rm const.}
\label{const}
\end{equation}
We expect a weak doping dependence of the left-hand side of the above equation, 
while 
\begin{equation}
T_c\propto E_F \propto n^{\rm (2D)}\propto 1/\lambda_{ab}^2(0), \; 
\mbox{ for } n \ll n_{\rm opt}
\end{equation}
can vary significantly upon going from underdoped to optimally doped 
regime $n_{\rm opt}$. In the overdoped regime $n> n_{\rm opt}$ the resistivity often displays non-linear temperature dependence. Simultaneously, if  the doping dependence of the effective mass is negligible for $n \ll n_{\rm opt},$ i.e. $m_{\rm eff}\approx {\rm const}$, the model predicts
\begin{equation}
\frac{{\rm d} \rho}{{\rm d} T} \propto 
\lambda_{ab}^5(0) \propto {1\over n^{5/2}}
\propto {1 \over T_c^{2.5}}, \;
\mbox{ for}\; n \ll n_{\rm opt}.
\label{proportionality}
\end{equation}
Let us note also that for a single-plane material ($d_0\equiv c_0$)  
only the 2D density $n^{\rm (2D)}=n c_0,$ i.e., the number of electrons per unit area, is relevant for the bulk 3D resistivity. The cancellation of the lattice constant $c_0$ can by easily understood inspecting the expression for the bulk conductivity of a system with equidistant conducting planes
$\sigma=c_0^{-1}n^{\rm (2D)}\tau_{\rm tr}(c_0)/m.$
According to the plane capacitor model, cf. Eq.~(\ref{Q2}), the scattering rate is 
proportional to $c_0^{-1}$ and, according to Eq.~(\ref{path}), the transport life-time $\tau_{\rm tr}\propto c_0.$
As a result, for single-plane materials $\sigma$ does not depend on the interplane distance $c_0,$ assuming $n^{\rm (2D)}={\rm const}.$
 
%------------------------------------------------------------------------------ 
\section{Numerical example} 
\label{sec:evaluation} 
%------------------------------------------------------------------------------ 
 
Let us provide now an estimate for $v_F$ and $l$ based on the proposed model for the set of parameters 
${\rm d}\rho/{\rm d}T = 0.5\;\mu\Omega {\rm cm/K}$~\cite{Rebolledo},
$n^{\rm (2D)}= {1\over2}a_0^{-2}= 3.37 \times 10^{14}\; {\rm cm^{-2}},$
$n =1 / (a_0^2 c_0) = 5.72\times 10^{21}\; {\rm cm}^{-3},$  
$2\pi\hbar/p_F=3.54 a_0,$
$a_0 = 3.85\; {\rm \AA},$ 
$c_0=11.8\;{\rm \AA},$ 
$d_0 = 3.18\; {\rm \AA},$ 
$m = m_{\rm eff} m_0,$  
$m_{\rm eff}=3,$~cf.~Ref.~\cite{Comment},
and
$m_0 = 9.11 \times 10^{-31}\; {\rm kg}$.
Substituting these parameters into Eq.~(\ref{rho}) and Eq.~(\ref{path}) we get an acceptable value for the Fermi velocity (cf. Table 3 of Ref.~\cite{Varlamov}, where 31, 140, 200 and 220 km/s estimations are cited) 
\begin{equation} 
v_F = \left({k_B \over 3 \pi^2 m} {a_0 c_0 \over d_0\varepsilon_0}  
{{\rm d} T \over {\rm d} \rho}\right)^{1/3} 
= 1.76\times 10^5 \; {\rm m \;  s^{-1}}= 176\mbox{ km/s}, 
\end{equation}  
and for the mean free path, respectively 
\begin{eqnarray} 
l(T=300\;{\rm K}) & = &5.67a_0=22\; {\rm \AA}, \\ 
\pi \left({l \over a_0}\right)^2 & \approx & 101. \nonumber 
\end{eqnarray}  
For $T_c=90\;{\rm K}$ 
and 
$\xi_{ab}(0)=12\; {\rm \AA}$ 
we have 
$\xi_{ab}(0)/l(T_c)= 16\% ,$
$\lambda_{ab}(0)=122$~nm,  $\lambda_{ab}(0)/ \xi_{ab}(0)=101,$
$8/(3 \sqrt{\pi})\approx 1.50,$ 
this numerical prefactor slightly changes if we apply more sophisticated approach for treatment of electric field fluctuations and the electron scattering by the random electric potential.
Finally for the constant $C_{\rho \lambda}$, introduced by Eq.~(\ref{const}) and describing the $\rho\,$-$\lambda$ correlations, we get
$C_{\rho \lambda}=320\; \Omega {\rm K}^{-1}(\mu {\rm m})^{-4}.$
These numerical estimates lead us to conclude that the suggested model does not contradict  the experimental data, cf.~Ref.\cite{Gurvich,Varlamov}.  
Furthermore, we consider that a detailed state-of-the-art derivation of the charge density fluctuation of the plasma in layered cuprates could be an adequate quantitative model for the theory of their electrical resistivity.  
Along the same line, a systematic study of the $\rho\,$-$\lambda$ correlations would provide an effective tool to analyse the scattering mechanisms in layered perovskites.
 
The mechanism of the electrical resistivity is qualitatively fairly simple and is schematically presented in Fig.~\ref{Scenario}. 
The conducting \cuo planes constitute plates of plane capacitors and one has to take into account the Boltzmann (or Rayleigh-Jeans) statistics of the electrostatic energy of the capacitors.  
The last criterion for applicability of the model is the significant low frequency reflection coefficient for an electromagnetic wave from a single \cuo plane.  
It exists only for high two-dimensional conductivity $\sigma c_0 > \varepsilon_0 c_{\rm light}$~\cite{Martin}, where $\rho_{ab} (4/c_0)= 300 \; \Omega$ sheet resistance is evaluated for Bi$_2$Ca$_2$SrCu$_2$O$_8$ just above $T_c.$ 
While moving in the conducting \cuo planes the charge 
carriers are scattered by charge density fluctuations in the same planes. 
In fact it is a self-consistent problem for collisionless plasma.  
The electrical resistivity appears upon taking into account the charge density fluctuations.  
This scattering mechanism is analogous to the Rayleigh's blue-sky law~\cite{Aslamasov,Landau8} where the light is scattered by fluctuations of the air density. Only the ``electron" sky is rather red. In case of Rutherford scattering the faster charges of smaller wavelength are scattered less intensively while for light the effect is opposite. 

%------------------------------------------------------------------------------ 
\section{Discussion and conclusions} 
\label{sec:conclusions} 
%------------------------------------------------------------------------------ 
 
The numerical example presented in Sec.~\ref{sec:evaluation} demonstrates that scattering by charge density fluctuations can explain the total resistivity of  layered cuprates or at least constitutes a significant part of it. The plane capacitor is an important structural detail of the scenario. 
Now we want to address two interesting issues. ({\it i}) Whether the existence of high-$T_c$ structures having linear resistivity is possible without the plane capacitor detail and vice versa? ({\it ii}) Why other layered structures do not display linear resistivity? The answers have qualitative character: 
 
 ({\it i}) High$-T_c$ superconductivity can exist even in artificial structures having a \cuo monolayer and the resistivity is yet linear. In this case the plane capacitors are missing; however, one has to take into account the electric field fluctuations close to the two dimensional (2D) \cuo layer. The 2D plasmons are gapless and overdamped above $T_c.$ Therefore we have to calculate 2D charge density fluctuations and the corresponding thermodynamic fluctuations of the electric field. Due to the equipartition theorem the linear resistivity is recovered again, with only the prefactor  being different. 
 
({\it ii}) There are various artificial metal-insulator structures where the temperature dependence of the in-plane conductivity is specific and non universal.  
When a metallic layer contains even a few monolayers we should take into account that the electric field does not penetrate through the metal. Thus due to trivial electrostatic reasons the thermodynamic fluctuations of the electric field in the insulator layers would be an inefficient mechanism for charge scattering in thick metallic layers. 
Metallic monolayers have residual resistivity related to defects, significant electron-phonon coupling etc. According to Mattissen's rule we could search for the charge density fluctuation part of the scattering, but it is unlikely that this mechanism dominates. 
The theory of fluctuations of the electromagnetic field between metallic layers 
(this is the geometry of  the Casimir effect) is a typical problem in  statistical physics, however, the difficult task will be the experimental separation of the linear term provided many other mechanisms contribute to the resistivity. 
 
To summarise, we came to the conclusion that an important hint for the applicability of the suggested model will be the existence of linear resistivity  
in other layered structures containing no \cuo planes but having  almost 2D charge carriers and of course good quality. If our simple electrostatic explanation is correct any layered material having good 2D metallic layers should display the same behaviour regardless of its electronic structure. 
In this case it is ensured that linear resistivity is not related to some specific subtle properties of electron band structure of  the \cuo plane or  
some sophisticated non-Fermi-liquid-like strongly correlated electron  
processes due to Cu$3d$ electrons, but rather to such a universal 
cause as the omnipresent fluctuations of the electromagnetic field  
and related to them charge density fluctuations---who could be blind for the blue sky. 
 
The linear resistivity of the layered ruthenates~\cite{Pavuna} could be considered as such an example and crucial experiment. 
In Sr$_2$RuO$_4$ the temperature dependence of the Hall coefficient is similar  
to the one measured in cuprates and the striking linear dependence of the conductivity persists over the whole temperature range 1--1000~K~\cite{Pavuna}. 
It is impressive to observe any physical quantity exhibiting linear behaviour 
over three orders of magnitude change of the temperature although it is a simple consequence of the conventional transport theory of metallic solids. 
The authors of Ref.~\cite{Pavuna} note that they were unaware of  
any model that specifically predicts or can convincingly account for  
essentially linear behaviour of resistivity over three decades of  
temperatures although most theories of high-$T_c$ superconductivity  
discuss linear resistivity. Nevertheless they suggest that linear resistivity is not an exclusive feature of the normal state of high-$T_c$ cuprates, but rather of all layered oxides especially perovskites, possibly even independently of the magnitude of $T_c$~\cite{Pavuna}. 
Indeed,  Sr$_2$RuO$_4$  ($T_c<1$K) presents an unequivocal demonstration that its linear resistivity is not related to processes involved in the pairing mechanism. 
The linear resistivity is created by thermally activated  electric fields while the pairing could originate in the conventional Heisenberg and Heitler-London type exchange attraction, cf. Ref.~\cite{MishonovBJP}. 
 
The emerging consensus that the model and mechanism of high-$T_c$ superconductivity must simultaneously explain both the linear normal resistivity and the superconducting gap is likely to be an erroneous presumption. It is not {\it a priori} clear why the high-$T_c$ superconductivity should repeat the history of phonon superconductors~\cite{Feynman}. One could think of the scattering mechanism as one that ``lives'' between the CuO$_2$ planes while the pairing mechanism is localized within them. 
The final solution of the problem, however,  would be rather given by the development of technologies for artificial layered structures necessary for the oxide electronics in the next millennium~\cite{Bozovic}.  
  
Concluding, we believe that there is nothing mysterious~\cite{Varlamov} in the linear dependence of the in-plane resistance in high-$T_c$ layered cuprates---it is just a consequence of the well-known physical laws dating back to the end of XIX and the beginning of XX century. 
\acknowledgments 
The authors are thankful to Prof.~J.~Indekeu and Prof.~F.~Vidal for the hospitality and fruitful discussions.  
The authors are very much indebted to Prof.~A.~Varlamov and Prof.~D.~Pavuna for the interesting correspondence and providing their results before publication. 
The authors appreciate also the interest of Prof.~V.~Moshchalkov to the present paper and his stimulating comments.  
One of the authors (TMM) is thankful to Prof.~A.~A.~Abrikosov,  Dr.~I.~Bozovic,
Dr.~D.~M.~Eagles, Prof.~A.~Leggett, Prof.~K.~Maki, and the referees, sending papers, correspondence, 
discussions and implied clarifications of the capacitor model of resistivity. 
Last but not least, special thanks to E.~Penev for critical reading of the manuscript, providing us with the electronic version of Fig.~\ref{Scenario} and many suggested improvements. 
This paper was partially supported by the Bulgarian NSF No.~627/1996, 
Visiting Professor fellowship from the Spanish MEC, 
The Belgian DWTC, the Flemish Government Programme VIS/97/01, the IUAP and the GOA.

%%% 
\begin{figure} 
  \begin{center}
  \epsfig{file=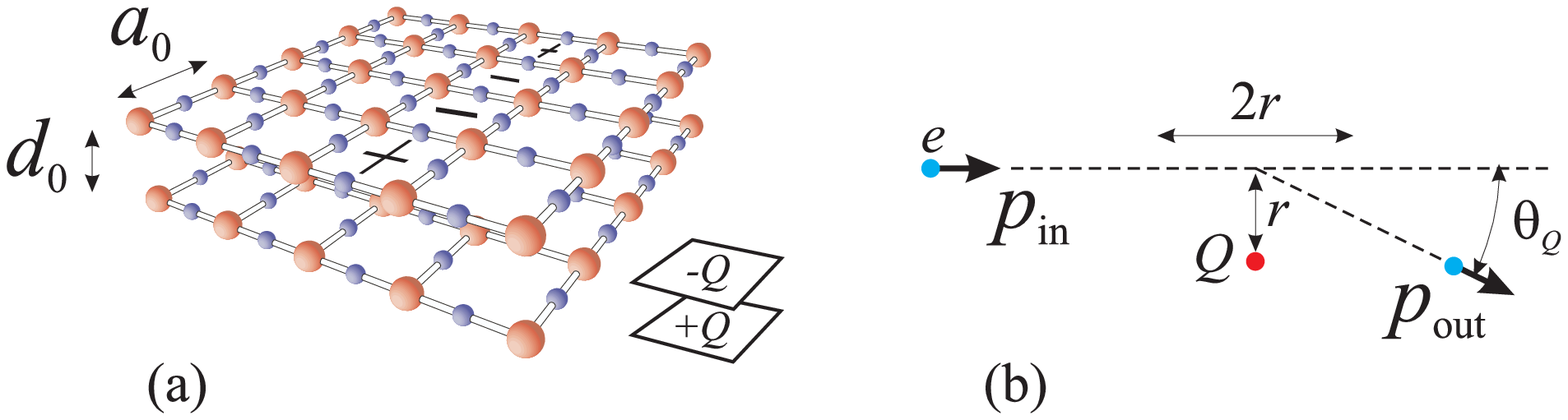,width=9.6cm}
  \vskip0.3cm
  \epsfig{file=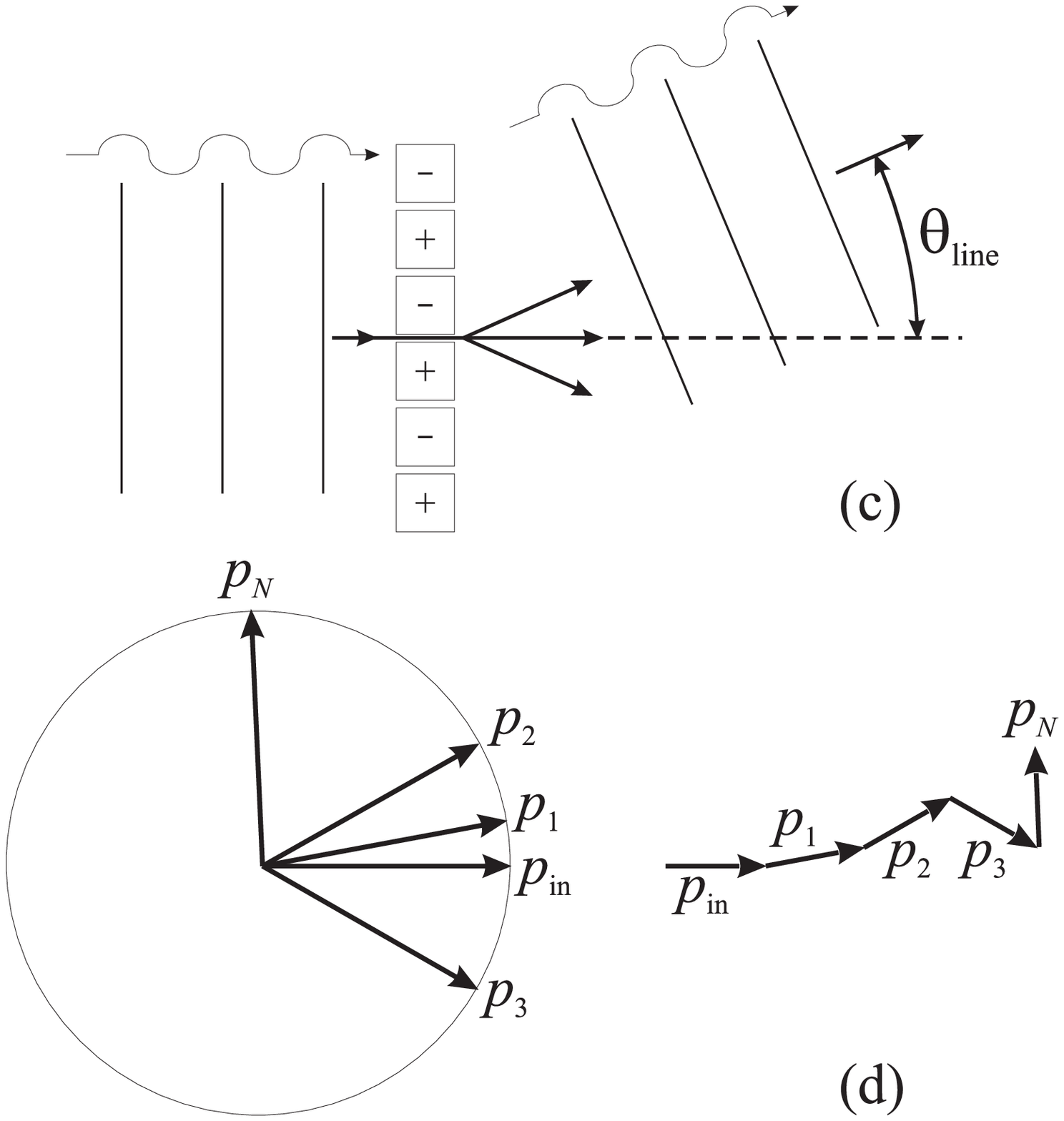,width=7.2cm}
  \end{center}
\caption{
The plane capacitor scenario for the resistivity of layered cuprates in a nutshell:
(a) Two conducting \cuo planes (spaced by distance $d_0$) of a layered perovskite are considered as an array of independent plain capacitors; ${1\over2}a_0$ is the Cu-O ion distance. In every capacitor we have thermally fluctuated charge $Q$ for the avaraged square of which the equipartition theorem gives $\langle Q^2\rangle/2C={1\over2}k_BT,$ where $C$
 is the capacitance;
(b) In the model evaluation the charges are considered as point ones. A charge carrier   
(a hole with charge $e$) with momentum $p_{\rm in}$ passes near the scattering centre $Q$ at distance $r$ and deviates at a small angle $\theta_Q(r)$, with $p_{\rm out}$ being its momentum after the scattering event;
(c) The electron waves are further scattered by a line of plane capacitors. 
The capacitor charges are independent random variables, thus the averaged square of the scattering  angle $\theta_{\rm line}$ is an additive quantity. The Rutherford cross section is the same in classical and quantum mechanics;
(d) Having crossed $N=l/a_0$ plaquete lines, the charge carrier "forgets" the direction   
of its initial momentum $p_{\rm in}$. This is employed to evaluate the mean free
 path $l.$ The diffusion of momentum on the Fermi surface is presented also as a trajectory   containing segments in the real space $\{p_{\rm in},p_1,\ldots,p_N\}.$
}
\label{Scenario}
\end{figure}
\end{document}